# Constraints on the symmetry energy and neutron skins from experiments and theory

M.B. Tsang[1], J. R. Stone[2], F. Camera[3], P. Danielewicz[1], S. Gandolfi[4], K. Hebeler[5], C. J. Horowitz[6], Jenny Lee[7], W.G. Lynch[1], Z. Kohley[1], R. Lemmon[8], P. Moller[4], T. Murakami[9], S. Riordan[10], X. Roca-Maza[11], F. Sammarruca[12], A. W. Steiner[13], I. Vidaña[14], S.J. Yennello[15]

[1]*National Superconducting Cyclotron Laboratory and Department of Physics and Astronomy, Michigan State University, East Lansing, Michigan 48824, USA*
[2] *Oxford Physics, University of Oxford, OX1 3PU Oxford, United Kingdom and Department of Physics and Astronomy, University of Tennessee, Knoxville, Tennessee 37996, USA*
[3]*CENTRA, Instituto Superior T´ecnico, Universidade T´ecnica de Lisboa,, Av. Rovisco Pais 1, 1049-001 Lisboa, Portugal*
[4]*Theoretical Division, Los Alamos National Laboratory, Los Alamos, New Mexico, 87545, USA*
[5]*Department of Physics, Ohio State University, Columbus, Ohio 43210, USA*
[6]*Center for Exploration of Energy and Matter and Physics Department, Indiana University, Bloomington, Indiana 47405, USA*
[7]*RIKEN Nishina Center, Hirosawa 2-1, Wako, Saitama 351-0198, Japan*
[8]*Nuclear Physics Group, STFC Daresbury Laboratory, Daresbury, Warrington, WA4 4AD, UK*
[9]*Department of Physics, Kyoto University, Kyoto 606-8502, Japan*
[10]*University of Massachusetts Amherst, Amherst, Massachusetts 01003, USA*
[11]*INFN, sezione di Milano, via Celoria 16, I-20133 Milano, Italy*
[12]*Physics Department, University of Idaho, Moscow, Idao 83844-0903, U.S.A*
[13]*Institute for Nuclear Theory, University of Washington, Seattle, Washington 98195, USA*
[14]*Centro de F´ısica Computacional, Department of Physics, University of Coimbra, PT-3004-516 Coimbra, Portugal*
[15]*Cyclotron Institute and Chemistry Department, Texas A&M University, College Station, Texas 77843, USA*

## Abstract

The symmetry energy contribution to the nuclear Equation of State (EoS) impacts various phenomena in nuclear astrophysics, nuclear structure, and nuclear reactions. Its determination is a key objective of contemporary nuclear physics with consequences for the understanding of dense matter within neutron stars. We examine the results of laboratory experiments that have provided initial constraints on the nuclear symmetry energy and its density dependence at and somewhat below normal nuclear matter density. Some of these constraints have been derived from properties of nuclei. Others have been derived from the nuclear response to electroweak and hadronic probes. We also examine the most frequently used theoretical models that predict the symmetry energy and its slope. By comparing existing constraints on the symmetry pressure to theories, we demonstrate how the contribution of the three-body force, an essential ingredient in neutron matter models, can be determined.

PACS numbers: 26.60.-c, 26.60.Kp, 21.65.Ef, 21.65.Cd,

## I. Introduction

Contemporary nuclear science aims to understand the properties of strongly interacting bulk matter at the nuclear, hadronic and partonic levels [1,2]. In addition to their intrinsic interest in fundamental physics, such studies have enormous resonance in astrophysics, from the evolution of the early universe to neutron star structure [3]. For example, a precise knowledge of the equation of state of neutron matter is essential to understand the physics of neutron stars and binary mergers, also predicted to be strong sources of gravitational waves [4]. While the equation of state of symmetric nuclear matter consisting of equal amount of neutrons and protons has been determined over a wide range of densities [5], our knowledge on asymmetric nuclear matter is very limited, largely as a consequence of our inadequate understanding of the symmetry energy [5,6]. The symmetry energy constrains the force on the number of protons and neutrons in a nuclear system. Its slope at saturation density provides the dominant baryonic contribution to the pressure in neutron stars [7]. It reduces the nuclear binding energy in nuclei and is critical for understanding properties of nuclei including the existence of rare isotopes with extreme proton to neutron ratios [8-10]. For all its importance, we do not have a realistic model of the nuclear force that can describe the equation of state of neutron matter [11-25]. Recently, substantial progress in our understanding of the symmetry energy has been made both experimentally [26-50] and theoretically [11-25,51,52], in particular at sub-saturation densities. This article will summarize the progress and indicate future avenues for such studies.

Using the Skyrme-Hartree-Fock model, Brown [6] showed a decade ago that selected Skyrme parameterizations, which fit the binding energy difference between $^{100}$Sn and $^{132}$Sn nuclei, may predict very different density dependencies of the energy per nucleon in pure neutron matter at densities above and below saturation density. Figure 1 shows that the symmetry energy, which governs the difference between the energies of symmetric and pure neutron matter, displays the same behavior. Brown also discovered a nearly linear correlation between the neutron skin thickness in heavy nuclei and the pressure of the neutron matter EoS at $\rho \approx 0.6\rho_0$, a trend replicated later by relativistic Hartree model calculations [53]. Many observables, from nuclear masses to nuclear structure and nuclear dynamics, also display significant sensitivities to the density dependence of the symmetry energy in the region near saturation density and below ($0.3 \leq \rho/\rho_0 \leq 1$). In this article, we will compare constraints derived from these observables.

To the lowest order expansion, the EoS of cold nuclear matter can be approximately written as the sum of the energy per nucleon of symmetric matter and an asymmetry term [5,51,52]

$$E(\rho,\delta) = E_0(\rho,\delta=0) + S(\rho)\delta^2, \qquad (1)$$

where $\delta = (\rho_n - \rho_p)/\rho$ is the asymmetry. $\rho_n$, $\rho_p$ and $\rho$ are the neutron, proton and nucleon densities, respectively.[i] $S(\rho)$ denotes the density dependence of the nuclear symmetry term $S$. It attains a value $S_o$ at normal (saturation) nuclear matter density, $\rho_o \sim 0.16$ nucleons·fm$^{-3}$, where the binding energy of symmetric nuclear matter reaches its maximum value of ~16 MeV.

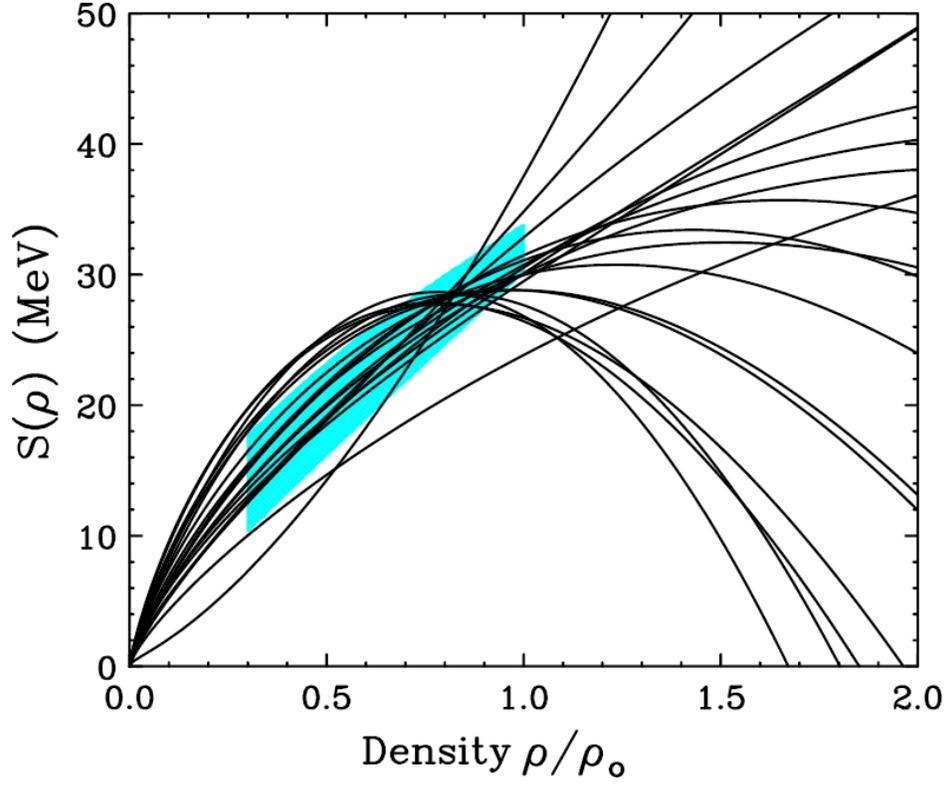

Figure 1: Density dependence of the symmetry energy from the Skyrme interactions used in Ref. [6]. The shaded region is obtained from heavy ion collisions experiments as described in the text and corresponds to the shaded region in Figure 2.

It is useful to expand the symmetry energy $S(\rho)$ in Equation (1) in a Taylor series around the saturation density, $\rho_o$,

$$S(\rho) = S_0 - L\varepsilon + K_{sym}\varepsilon^2 + O[\varepsilon^3], \tag{2}$$

where $\varepsilon \equiv (\rho_o - \rho)/3\rho_o$, $L$ and $K_{sym}$ are the slope and curvature parameters at $\rho_o$.[ii] The slope parameter $L$,

$$L = 3\rho_0 dS(\rho)/d\rho|_{\rho_0} = [3/\rho_0]P_0, \tag{3}$$

governs $P_0$, the pressure from the symmetry energy in pure neutron matter at $\rho_o$. $P_o$, provides the dominant baryonic contribution to the pressure in neutron stars at $\rho_o$ [7] and influences the inner crusts and radii of neutron stars [3, 54]. Thus $L$ forms an essential link between nuclear physics and astrophysics [7].

Realistic models of nuclear matter and its effective interactions predict model dependent correlations between $S_o$, $L$ and $K_{sym}$. Observables with different sensitivities to $S_o$, $L$ and $K_{sym}$, can be combined to allow independent constraints on them and on the theories from which they can be calculated. However, $K_{sym}$ correlates strongly with $L$, and contributes weakly to the symmetry energy at sub saturation densities, making it difficult to constrain $K_{sym}$ [55]. In the following, we discuss constraints on $S_o$ and $L$ extracted from ground state properties, such as nuclear masses and neutron skin thicknesses and from excited state properties, such as the energies of isobaric analog states, the energies and strengths of giant and pygmy dipole resonances. We also discuss constraints provided by observables sensitive to the transport of neutrons and protons during nucleus-nucleus collisions. In addition, we compile results from recent measurements of the neutron skin thickness of $^{208}$Pb, the best candidate for precise neutron skin measurement. We examine critically the consistency between different measurements including experimental and theoretical uncertainties and discuss the ability of future experiments to provide further constraints.

We note that these constraints on $S_o$ and L are most directly applicable to matter at uniform or nearly uniform density. Low density $\rho<0.3\rho_0$ matter plays important roles in neutron stars and core collapse supernovae and is not uniform [56], Consequently, the mean field does not directly apply. Laboratory experiments have been performed to investigate the properties of low-density nuclear matter. We refer readers to the work of Natowitz et al. [57] for a recent exploration of the interplay of clusterization and the low-density symmetry energy in laboratory systems. The constraints on $S_o$ and L discussed in this paper can be relevant to statistical models that describe the separation of matter into dilute and dense phases, where the latter resemble nuclei [58].

This paper is organized as follows. Recent experimental measurements providing constraints on the nuclear symmetry energy and the neutron skin thickness of $^{208}$Pb are discussed in Section II. Discussion of the effects of the constraints on different theoretical models is presented in Section III, followed by summary and outlook in Section IV.

## II. Experimental constraints on the symmetry energy

### A. Symmetry energy constraints from nucleus-nucleus collisions

Large density variations can be attained for a very short period of time in Nucleus-Nucleus Collisions using heavy projectiles (HIC) such as Au or Sn [5, 51, 52]. The EoS is an essential input to the transport models, and can be constrained by comparing measurements of such collisions to transport model calculations [5, 51, 52, 59, 60]. This strategy was successfully applied to constrain the EoS of symmetric matter, $E_0(\rho, \delta=0)$, at densities of $\rho_0 \leq \rho \leq 5\rho_0$, by studying energetic $^{197}$Au+$^{197}$Au collisions [5, 27, 59].

To gain sensitivity to the symmetry energy and extrapolate to neutron-rich asymmetric matter, one can vary the neutron and proton numbers of the projectile and target nuclei so as to compare emission of particles from neutron-rich systems to that from neutron-deficient systems [28-36]. The influence of the symmetry potential can be more easily distinguished from other effects by comparing the emitted nucleons and light nuclei with different neutron and proton numbers. Especially interesting are the comparisons between "mirror" nuclear ejectiles with the same mass, and total isospin, but where the proton and neutron numbers are exchanged, e.g. comparing the emission of neutrons to protons, $^3$H to $^3$He, $^7$Li to $^7$Be [27,30,32,51]. Such comparisons probe the combination of Coulomb and symmetry mean field potentials. The latter has the opposite sign for protons and neutrons but the combination contributes greatly to the uncertainty in the symmetry energy [51].

In a neutron rich environment, the symmetry potential tends to expel the neutrons and attract protons, enhancing the yield ratios of ejected neutrons/protons and other isotopes and influencing their dependence on the ejected particle's momentum [30,51,52]. Neutron-proton spectral ratios [30], and neutron, hydrogen [31] and fragment flows [32, 33] have all been used to study the density dependence of the symmetry energy. When the collision involves projectiles and targets of different N/Z asymmetry, and different local density, $\delta = (\rho_n - \rho_p)/\rho$, the symmetry potential pushes the system towards an "isospin equilibrium" characterized by constant values for $\delta$ throughout the system. Thus, the magnitude of the symmetry potential in a low-density "neck" region joining a projectile and target during a peripheral or mid-central collision governs the rate of "isospin diffusion" between a projectile and a target of different asymmetry. This phenomenon has been used to probe the symmetry energy [28, 29, 34-36].

The neutron and proton spectra from central collisions for $^{124}$Sn+$^{124}$Sn and $^{112}$Sn+$^{112}$Sn collisions at 50 MeV per nucleon [30] have been measured. At the same incident energy, isospin diffusion was investigated. Normalization of the latter requires asymmetric systems $^{124}$Sn+$^{112}$Sn, ($^{112}$Sn+$^{124}$Sn,) to be compared to symmetric systems of $^{124}$Sn+$^{124}$Sn and $^{112}$Sn+$^{112}$Sn collisions [29, 35, 36]. A chi-square analysis compared measured and calculated values for the ratios of neutron to proton spectra as well as two observables sensitive to isospin diffusion [26]. In Figure 2, a set of constraints corresponding to 2 standard deviations from the minimum, corresponding to 95% confidence levels, is shown as a shaded band bounded by two diagonal lines in the (L, $S_0$) plane. The solid star shows the results from isospin diffusion observables measured for collisions at a lower incident energy of 35 MeV per nucleon [36]. The corresponding density dependence of the symmetry energy are plotted as the shaded region in Figure 1. All observables obtained from the sets of data described here were mainly sensitive to the symmetry energy at densities around half of the normal density, $\rho \approx 0.5\rho_0$.

Transverse collective flows of hydrogen and helium isotopes [32,33] as well as intermediate mass fragments with $Z < 9$, have also been measured at incident energy at 35 MeV per nucleon in $^{70}$Zn+$^{70}$Zn, $^{64}$Zn+$^{64}$Zn and $^{64}$Ni+$^{64}$Ni collisions and compared to transport calculations. These comparisons yielded values for $(S_0, L)$ denoted by the open squares in Figure 2. No extensive chi-square or sensitivity analyses have been performed over the $(S_0, L)$ space for either the transverse flow data (open squares) or the low energy isospin diffusion data (solid star). We consequently plot them without error bars.

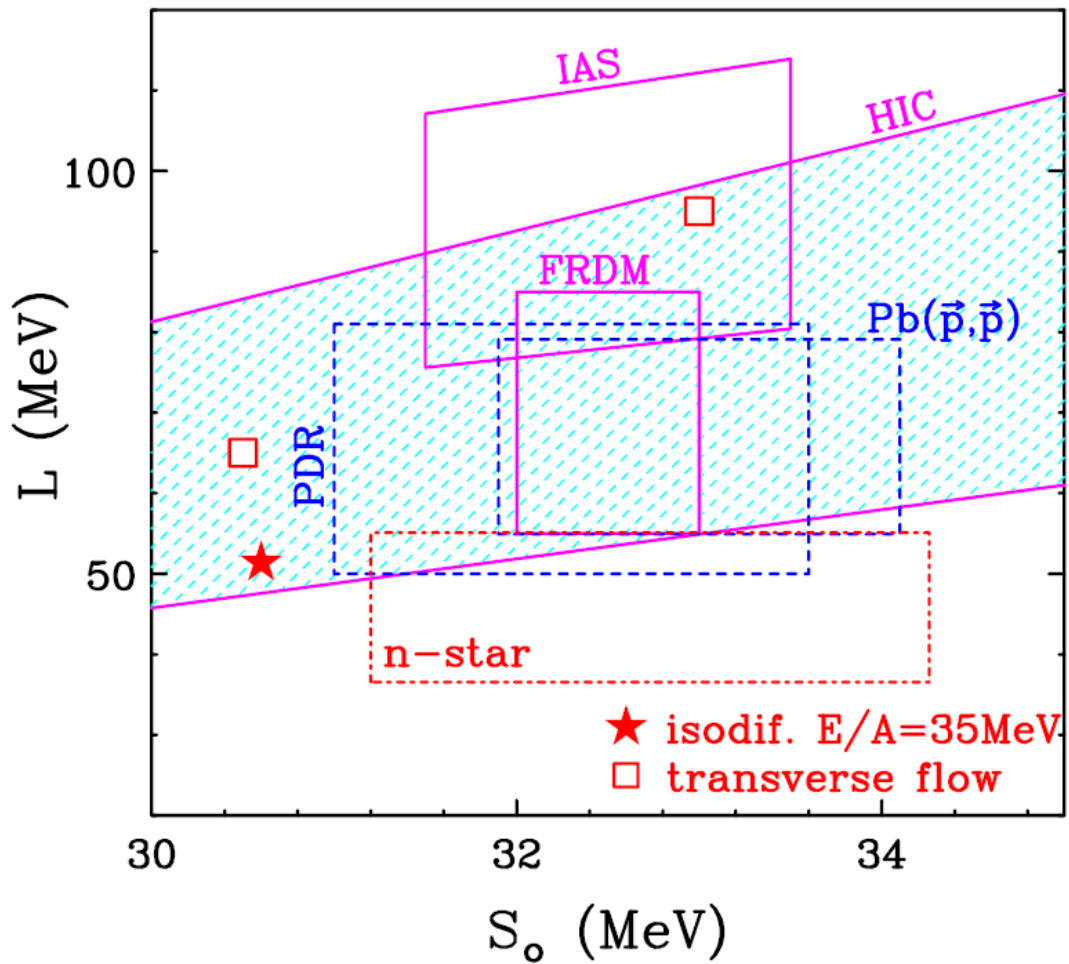

Figure 2: (color online) Constraints on the slope $L$ and magnitude $S_0$ of the symmetry energy at saturation density from different experiments. The experimental methods are labeled next to the boxes with the estimated uncertainties. The symbols are results without the analysis of the errors. See text for details.

## B. Symmetry energy constraints from nuclear binding energies

In 1935, Bethe and Weizäcker proposed a very successful theory of nuclear binding energies [8,9]. To a good approximation, this theory reduces to a remarkably simple mass formula in which the binding energy B(N,Z) is obtained as a function of proton number Z and neutron number N, with mass number, A=N+Z

$$B(N,Z) = a_{vol}A - a_{surf}A^{2/3} - a_C \frac{Z^2}{A^{1/3}} - a_{sym}(A)\frac{(N-Z)^2}{A} + dE ,  \quad (4)$$

in terms of volume, surface, Coulomb, symmetry energies, and additional small contributions, related to microscopic effects [8,9]. The coefficients in Eq. (4) can be determined by fitting to known atomic masses across the nuclear chart.

Myers and Swiatecki separated the volume and surface contributions to the symmetry energy in their liquid drop model [61] and subsequently developed a refined version of the model, called the droplet model, by expanding the volume, surface and Coulomb energies in a Taylor series in terms of (N-Z)/A and $A^{-1/3}$ around the standard liquid drop model values. This introduced additional degrees of freedom allowing for the deviations from uniformity of the proton and neutron densities and led to a more realistic parameterization of the symmetry energy [61].

Since the symmetry energy contribution to the total binding energy can be small relative to those from the other terms, the unambiguous determination of, $S_0$ and $L$, the magnitude and slope of the symmetry energy at saturation density, has proven difficult [54, 62]. To overcome the problem, all main contributions to the binding energy must be theoretically described with highest possible accuracy. A refinement of the droplet model, the finite-range droplet model (FRDM) [10], came close to fulfillment of this requirement. It included additional important features such as microscopic "shell" effects and the extra binding associated with N = Z nuclei. The FRDM reproduced binding energies of known nuclei with a deviation σ = 0.67 MeV and implied a value of $S_0$ = 32.73 MeV [10].

Despite the greatly improved predictions for nuclear binding energies it provided, it still did not have enough sensitivity to determine $L$. More complex calculations [10] that include additional effects such as axially asymmetric nuclear ground state shapes, further improved the deviation of nuclear binding energies to $\sigma = 0.57$ MeV. This means that the nuclear binding energies are reproduced to within 0.1%. The model now allows determination of both $S_0 = 32.5 \pm 0.5$ MeV and $L = 70 \pm 15$ MeV. This constraint is shown as a square box in Figure 2, labeled "FRDM" on the top. Although the results are consistent with other model predictions, this very small uncertainty in the value of $S_0$, if correct, limits seriously the choice of currently available equations of state used in modeling neutron stars and supernova matter.

### C. Symmetry energy constraints from Isobaric Analog State energies

Fits of nuclear binding energies to mass models, must address ambiguities stemming from the similarities in the influences of Coulomb and symmetry energy terms over the range of experimentally measured masses. These ambiguities in the determination of the symmetry energy from binding energies caused by the Coulomb term can be removed by taking advantage of the charge independence of nuclear interactions, i.e. to a very good approximation, strong interactions between nucleons in the same state do not depend on whether the nucleons are protons or neutrons [38,55]. For example, there is an excited state in the nucleus $^{12}$C (Z=N=6) with the same wavefunction and nuclear contribution to the binding energy as the ground state of its isobar, $^{12}$B (Z=5,N=7). This $^{12}$C excited state is called the "isobaric analog" of the ground state of $^{12}$B and its binding energy differs from that of the $^{12}$B ground state by the difference in their Coulomb energies. Similarly, there is a higher lying excited state in $^{12}$C that is the isobaric analog of the ground state of $^{12}$Be and both states differ only in their Coulomb energies. It follows from Eq. (4) that the energy differences between the ground state for a nucleus with N>Z and the isobaric analogs of the ground states of neighboring isobars are given by the symmetry energy.

Many such states called the Isobaric Analog states (IAS), have been identified [63]. To utilize this technique, however, one must realize mathematically that the nuclear Hamiltonian is charge independent to a good approximation. It depends on the total isospin operator $T^2$ and not on its projection $T_z$. which has the eigenvalue (N-Z)/2 for a nucleus with neutron and proton number N and Z, respectively. This can be accomplished by replacing $E_{sym} \approx a_{sym}(A) \frac{(N-Z)^2}{A}$ with $E_{sym} = \frac{4a_{sym}(A)}{A} T(T+1)$ [38]. This allows the asymmetry coefficient, $a_{sym}(A)$, to be determined on a nucleus-by-nucleus basis from

$$a_{sym}(A) \approx -\frac{\Delta B}{4\Delta T^2} A, \quad (5)$$

where $\Delta T^2 = T_{IAS}(T_{IAS}+1) - T_{gs}(T_{gs}+1)$ is the difference between the known $T^2$ eigenvalues for the Isobaric Analog states and the ground state in the same nucleus, and $\Delta B$ is the differences in the binding energies of these two states. By fitting the available data on the IAS, Danielewicz and Lee obtained the constraint [38, 55] shown as a parallelogram in Figure 2, labeled "IAS" inside the box. Further refinements to these fits are in progress [38].

### D. Neutron skin thickness measurements

In light nuclei with $N \approx Z$, the neutrons and protons have similar density distributions. With increasing neutron number, the radius of the neutron density distribution becomes larger than that of the protons, reflecting the pressure of the symmetry energy. The difference $\delta R_{np}$ in the neutron and proton root mean square radii is called the neutron skin, i.e.

$$\delta R_{np} = \langle r^2 \rangle_n^{1/2} - \langle r^2 \rangle_p^{1/2}. \quad (6)$$

Proton radii have been determined accurately for many nuclei using electron scattering experiments [64]. This accuracy reflects the accuracy of perturbative treatments of the electromagnetic process. The neutron density distribution is more difficult to measure accurately because it interacts mainly with hadronic probes (protons, alphas, pion and antiprotons), through non-perturbative interactions, the theoretical description of which is model dependent.

The stable nucleus $^{208}$Pb is a very interesting candidate for determining the neutron skin. With a closed neutron shell at N=126 and a closed proton shell at Z=82, it is very asymmetric. The closed shells make its structure relatively well understood, allowing a clean relationship between skin thickness and properties of the symmetry energy. The relationship between the neutron skin thickness of ~0.2 fm and the radius of neutron star of ~12 km, predicted by many models [6,7,53,54], has stimulated many experiments to measure the neutron radius of $^{208}$Pb [39-46]. We note, however, that the neutron star radius reflects the pressure due to the symmetry energy at a range of densities and is also highly sensitive to its pressure at 2-3 times saturation density [3].

In the following subsections, we discuss experimental measurements that probe neutron skins using both electroweak and hadronic probes. These experiments require models to extract the neutron skin thickness of $^{208}$Pb from the data. Unfortunately, not every published value for the skin thickness includes an estimate of the theoretical uncertainty; Following Brown, we estimate such uncertainties to be of the order of 0.05 fm [65]. We therefore adopt a minimum error in $\delta R_{np}$ of ±0.05 fm in order not to bias comparisons of various $\delta R_{np}$ values towards values with underestimated uncertainties.

1.     **Parity Violating Electron Scattering.**

The possibility of measurements of the neutron radius in $^{208}$Pb by the **P**b **R**adius **EX**periment (PREX) at Jefferson Laboratory had been widely discussed [66,67]. This experiment is designed to extract the neutron radius in $^{208}$Pb from parity violating electron scattering. The electroweak probe has the advantage over experiments using hadronic probes in that it allows a nearly model independent extraction of the neutron radius that is independent of most strong interaction uncertainties [66]. The experimental signature, however, is very small, making high precision measurements very challenging [67]. Due to technical problems in a recent measurement [39], which reduced the statistics severely, the extracted $^{208}$Pb skin thickness had a large statistical uncertainties of $\delta R_{np}$=0.33(+0.16,-0.18) [39]. A second experimental run has now been approved for PREX to run in a few years to achieve the original goal of measuring the skin thickness in $^{208}$Pb with an uncertainty of ± 0.05 fm in $\delta R_{np}$ [68].

2.  **Proton elastic scattering of $^{208}$Pb ($\vec{p},\vec{p}$)**

Zenihiro et al. recently reported an extraction of the neutron skin thickness of $^{208}$Pb via polarized proton elastic scattering [40]. Cross-sections and vector analyzing powers for polarized proton elastic scattering on $^{58}$Ni and $^{204,206,208}$Pb were measured with high precision. A t-matrix parameterization of the proton optical potential was constrained by the $^{58}$Ni measurements, and fit to the $^{204,206,208}$Pb data by adjusting the neutron densities. The deduced neutron skin thickness of $^{208}$Pb, $\delta R_{np}$ = 0.211 (+0.054, -0.063) fm is plotted as a solid circle in the inset in Figure 4. This uncertainty includes an estimated theoretical uncertainty arising from using different models for the nucleon-nucleon interaction. The symmetry energy constraints consistent with the determined skin thickness of $^{208}$Pb were evaluated within both relativistic and non-relativistic models, and plotted as the dashed blue rectangular box labeled "Pb($\vec{p},\vec{p}$)" in Figure 2.

In addition to the case of doubly magic $^{208}$Pb, we note that proton scattering and interaction cross section measurements have probed neutron skins in other nuclei such as Sn [41] and Na isotopes [42], respectively.

3.  **Antiprotonic Atoms**

Neutron skins of many other nuclei have been probed by measurements [43, 44] of photons emitted during the decays of anti-protonic states of high orbital angular momentum where the angular momentum barrier restricts the interactions between the anti-proton and the neutron density to large distances. The root mean square neutron radius is not directly measured, making the skin thickness results strongly dependent on theoretical models. Nevertheless, systematic measurements of the neutron skins of a range of nuclei can contribute significantly to a global understanding of the evolution of neutron density distributions with nuclear charge and mass. A systematic analysis of the neutron skin of 26 antiprotonic atoms ranging from $^{40}$Ca to $^{238}$U [43, 45] suggests that there is a correlation between the neutron skin thickness and the isospin asymmetry. The adopted value of $\delta R_{np}$=0.18±(0.04)$_{exp}$±(0.05)$_{theo}$ fm from averaging the results of [44, 45, 67] is shown as an open circle in Figure 4. In Ref. [45] a Droplet Model is used to determine a correlation between $\delta R_{np}$ and $L$ resulting in the constraints, $L$ = 55 ± 25 MeV and $S_o = 31.5 \pm 2$ MeV. The uncertainties of the results from the antiproton experiments are larger than those in the other experiments discussed here. Nonetheless, the method of using the systematics of a range of nuclei to extract the skin thicknesses of $^{208}$Pb remains attractive especially if both the experimental and theoretical uncertainties could be reduced when good quality skin thickness measurements of a wide range of nuclei become available.

## 4. Electric Dipole Strength Function

### a. Electric Dipole Polarizability

The Electric Dipole Polarizability (EDP) $\alpha$, is defined by the relationship $\vec{p} = \alpha \vec{E}$ between the induced electric dipole moment $\vec{p}$ of a nucleus and the static electric field $\vec{E}$ that induces it. While the polarizability of some light nuclei has been determined by placing them in the field of a heavy target nucleus [69], it is more common to excite the nucleus by photo-excitation [70,71] or Coulomb excitation [46] and use the relationship,

$\alpha = \dfrac{\hbar c}{2\pi^2 e^2} \int \dfrac{\sigma_{ABS}}{\omega^2} d\omega$, between $\alpha$ and the photon absorption cross section $\sigma_{ABS}$ weighted with $\omega^{-2}$ and integrated over the incident photon energy, $\omega$.

The interaction of nuclei with an electric dipole field leads to the real or virtual excitation of nuclear excited states, many of which contribute to the Giant Dipole Resonance (GDR) in nuclei [47]. Semiclassically, the GDR can be viewed as a collective vibration in which neutrons and protons move in opposite directions, displacing the neutron and proton densities relative to each other and increasing the symmetry energy. Thus, the symmetry energy contributes to the "restoring force" for the vibration, and strongly influences the excitation energies of states that can be easily excited by an electric field [71,73].

In neutron rich nuclei a significant enhancement in the EDP can be expected due to the development of a neutron skin. In such nuclei, one has the possibility of vibrations of the (N,Z) symmetric core against the neutrons in the skin. This leads to the appearance of low energy states that are easily excited by electric dipole radiation, i.e. low-energy dipole strength, which greatly enhances $\alpha$ due to the $\omega^{-2}$ weighting in the integral above. In fact, the low energy dipole strength can contribute as much as 25% to the dipole polarizability [74]. For very neutron-rich nuclei, this enhanced dipole strength can be localized in energy and appears in the form of a "pygmy dipole resonance (PDR)" discussed in the next section.

Reinhard and Nazarewicz predicted a strong correlation between the neutron skin thickness in $^{208}$Pb and the EDP within a model with a Skyrme interaction and an effective Lagrangian [75]. Recent experiments using inelastic scattering of polarized protons on $^{208}$Pb at very forward angles [46] have provided the complete electric dipole response of $^{208}$Pb from low excitation energies up to the giant dipole resonance (GDR) with high resolution. A precise value of the EDP was extracted and the calculations from Ref. [75] were used to predict a value for the neutron skin thickness $\delta R_{np}$ = 0.156 (+0.025, -0.021) fm for $^{208}$Pb [46]. More recent calculations have shown the skin thickness results to be somewhat model dependent [76]. Accordingly, the predicted value for $\delta R_{np}$ from [46], is shown in the inset of Figure 4 as an open blue square with a larger uncertainty of ±0.05 fm.

b. **Pygmy Dipole Resonances.**

In very neutron-rich nuclei such as $^{68}$Ni and $^{132}$Sn, enhanced low energy electric dipole strength has been observed and attributed to a Pigmy Dipole Resonance that peaks at excitation energies well below the GDR [48-50]. These PDR peaks can exhaust in the order of 5% of the energy-weighted sum rule (EWSR). In many models, the calculated percentage of the energy-weighted sum rule (EWSR) exhausted by PDR is shown to be linearly related to the slope parameter, $L$, of the symmetry energy. Carbone et al. extracted a value of $L$ = 64.8 ± 15.7 MeV [49] from measurements of the PDR for $^{68}$Ni [50] and $^{132}$Sn [48]. In addition, they utilized the correlations between $L$ and $\delta R_{np}$ within various models to predict the skin thickness for $^{68}$Ni and $^{132}$Sn and extrapolate $\delta R_{np}$ = 0.194 ± 0.024 fm for $^{208}$Pb. The latter value is plotted as a solid square in Figure 4. To be uniform with our concerns about model dependencies in the extracted values for $\delta R_{np}$, we adopt an uncertainty of ±0.05 fm. In addition, the authors [49] obtain a value of $S_o$ =

32.3±1.3 MeV using the correlation between $L$ and $S_0$, calculated by the same models. The PDR symmetry energy constraint is shown as a dashed rectangle in Figure 2 with the label "PDR" at the bottom of the box.

### III.  Theoretical Models of Nuclei and Nucleonic Matter

In this section, we discuss calculations of the symmetry energy using representative theoretical models. We examine the range of calculated symmetry energies and neutron skin thickness predicted by calculations using phenomenological interactions and by calculations using microscopic and phenomenological two nucleon forces and phenomenological three- and more- body forces.

**A. Phenomenological models**

Many calculations employ effective, density dependent nucleon-nucleon interaction of a Skyrme or Gogny type [11-13] or meson exchange interactions based on a relativistic mean-field model (RMF) Lagrangian approach [14,15]. The strength and ranges of the various terms in these phenomenological interactions are adjusted to describe nuclear properties, with little direct input from nucleon-nucleon scattering.

The interactions utilized by these Skyrme, or RMF approaches, typically have a number of adjustable correlated parameters, which have not been adequately constrained by existing data, leading to a proliferation of different parameterizations. Figure 3 shows the $S_0$ - $L$ correlation predicted by a selection of these phenomenological parameterizations. The open circles show predictions of the $S_0$ - $L$ correlation for a set of Skyrme interactions used in Ref. [62]. The range of possible $S_0$ - $L$ correlations spanned by these interactions extends well beyond the experimental constraints discussed previously. This range is typical of predictions of Skyrme interactions.

One of the ultimate goals for studies of the symmetry energy should be to narrow the experimentally and theoretically allowed region in the $S_0 – L$ plane. To illustrate such an exercise, Dutra et al. [16] extended the set of constraints, usually utilized in development of Skyrme parameterization, to 11 macroscopic conditions, originated from empirical properties of nuclear matter and experiments and 4 microscopic constraints including density dependence of the effective mass and Landau parameters, to test the suitability of 240 Skyrme interactions. The combined effect of these constraints leaves only 5 Skyrme interactions, that satisfied nearly all the constraints. The L and $S_0$ values calculated with the selected interactions (solid circles) cluster along the lower boundaries of the experimentally allowed regions in Fig. 3.

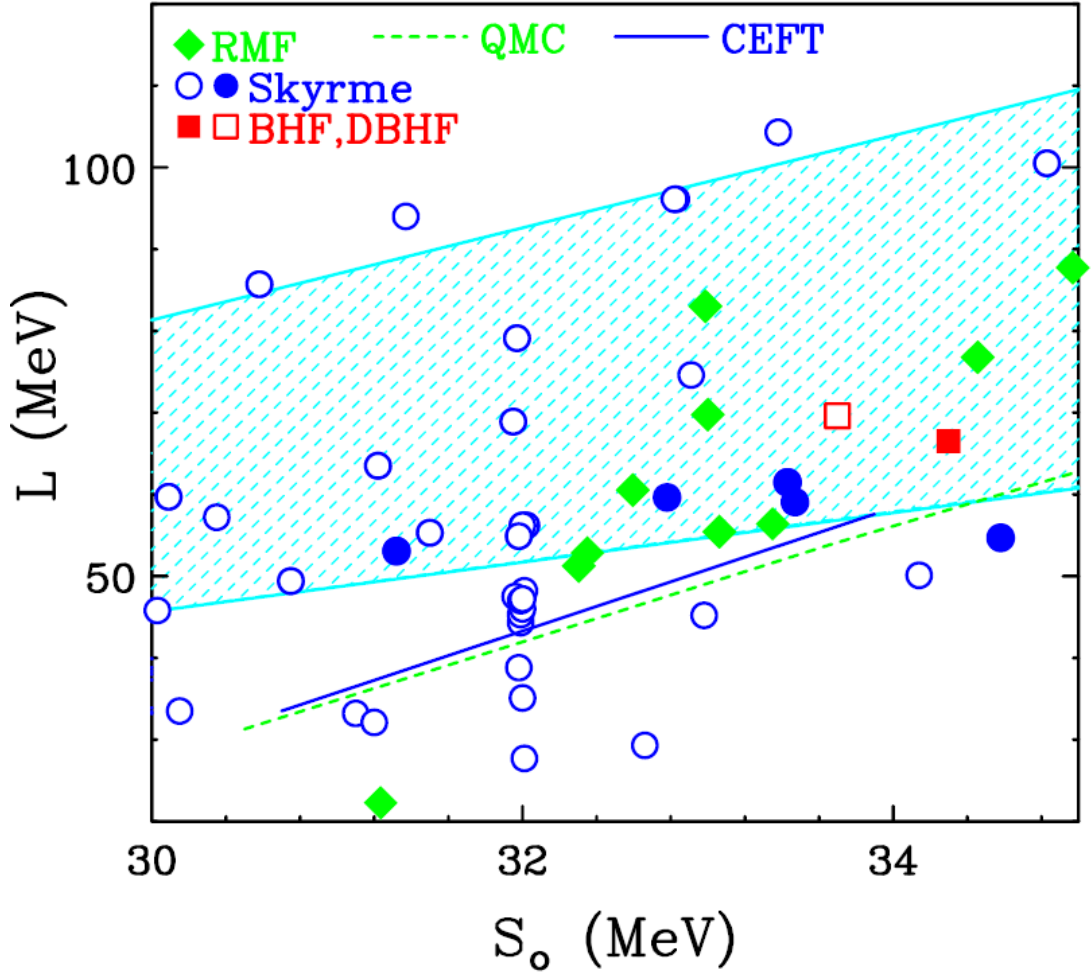

Figure 3: (color online) Symmetry energy correlations from different models (symbols). The dashed and solid lines represent the linear relationship between L and $S_O$ in the QMC and CEFT models respectively. The shaded region represents constraints obtained from heavy ion collisions experiment. The axes are the same as Figure 2.

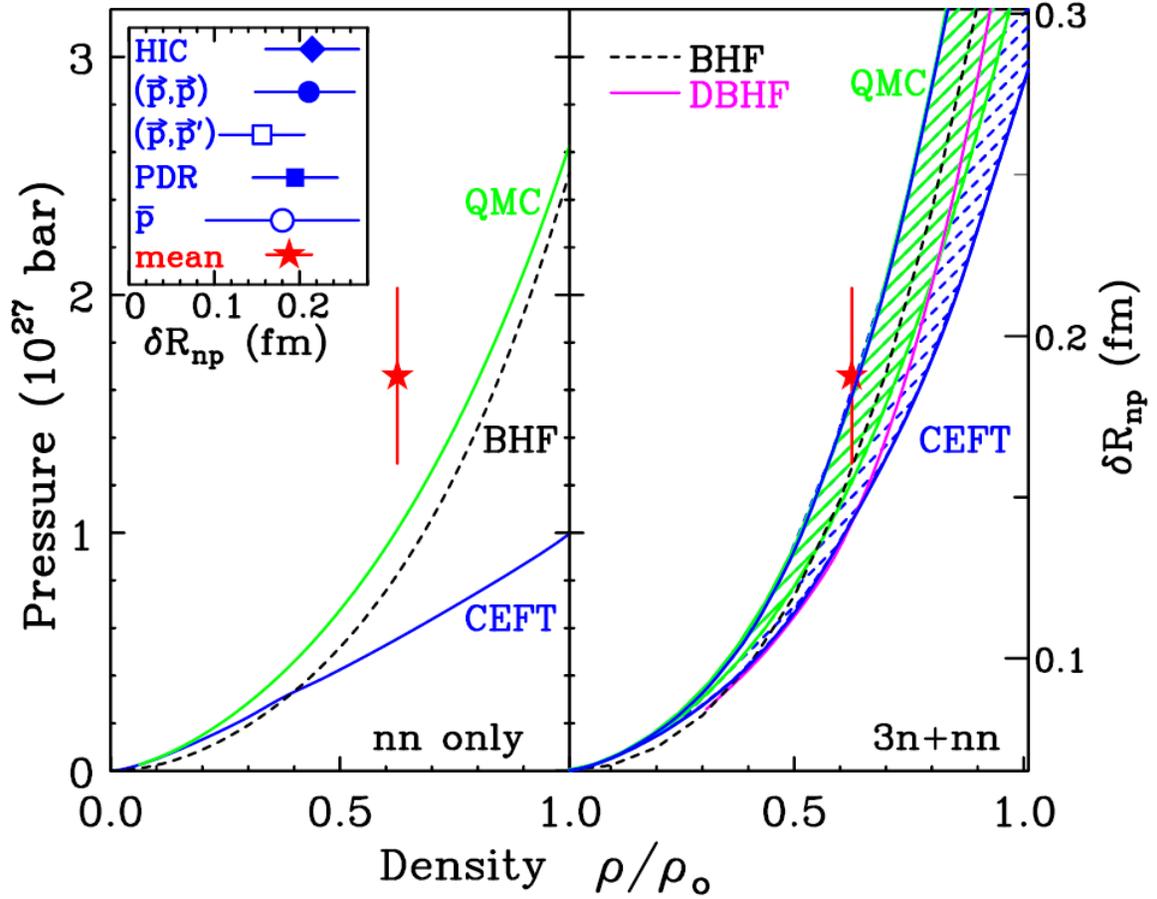

Figure 4: (color online) Density dependence of pressure in pure neutron matter as predicted in BHF, QMC and CEFT models without (left panel) and with (right panel) 3-body neutron forces. The inset in the left panel shows experimental data currently available on the neutron skin thickness in $^{208}$Pb and the red star indicated their weighted average. For more details see text.

**B. Microscopic models with free interactions**

Most microscopic approaches start from free two-body nucleon-nucleon interactions (NN) that reproduce nucleon-nucleon scattering and three-body interactions, which together with the two-body interactions, reproduce bound state properties of selected light nuclei [17,18]. The in-medium correlations to these interactions can then be calculated by many-body techniques, such as the Brueckner-Hartree-Fock (BHF) or its relativistic counterpart Dirac-Brueckner-Hartree-Fock (DBHF) (see e.g. [19,20]). We note that most of the bulk of the relativistic effects typical of DBHF approaches can be associated with the class of three-body forces originating from virtual pair excitation.

The Quantum Monte Carlo (QMC) technique provides another way to include the many body correlations and solve the many-body problem. This technique can tackle the problem exactly when the interactions are local in configuration space. However, it requires significant computational resources, making large nuclei impossible at present to compute. QMC calculations for nuclear matter have successfully demonstrated a strong correlation between the symmetry energy and its density dependence [21,22] shown as the dash line in Figure 3.

We note that this strong correlation has been recently combined with the constraints on the mass and radius of neutron stars to give $31.2 < E_{sym} < 34.3$ MeV and $36 < L < 55$ MeV to 95% confidence level [77]. These constraints, derived from theoretical analyses of x-ray data from satellite observatories, are shown as a red dashed box labeled as n-star in Figure 2. The constraints tend to favor a weak density dependence of symmetry energy, barely overlapping with the lower bound of the boundaries determined from nuclear experiments. We note that the small extracted values of the radii of neutron stars depend on assumptions regarding the dynamics of x-ray bursts and the emissivity of the stellar surface [78]. It will be an important scientific objective that both laboratory measurements and astrophysical observations can be described consistently with the same assumptions about the density dependence of the symmetry energy.

Finally, chiral effective field theory (CEFT), with renormalization group (RG)-evolved interactions constrained by nucleon scattering data [23], has recently been used to calculate nuclear matter properties [24, 25]. These calculations constrained the pressure of neutron matter at saturation densities within ±25%.

Figure 4 illustrates theoretical predictions of the density dependence of the pressure in pure neutron matter. The left panel shows results that include 2-body potentials without inclusion of 3 body neutron (3n) forces. The curves are calculated using BHF approach with the Av18 potential [19], the QMC approach with the Av8' potential [22], and the CEFT approach. In the right panel we demonstrate the effect of including 3n forces and show results using the BHF approach with the Av18 + UIX 3-body potential [19], the DBHF approach with the relativistic one-boson-exchange Bonn B potential [20], the QMC approach with the $Av8' + V_{2\pi}^{PW} + V_{m=150}^{R}$ potential (lower limit) and the Av8' + UIX potential (upper limit of the green shaded area) [22]. These limits are derived from the spread of predictions, calculated using different forms of the 3-neutron force [22]. The lower and upper limit of the CEFT predictions, reflect the theoretical error of the model [24, 25]. Even though the calculations shown in the right panel of Figure 4 have been calculated with different models using different forms of 3-body potentials, all calculations lie within the uncertainties of the CEFT calculations (blue shaded region). Furthermore, the upper limits obtained for the QMC and CEFT approaches are almost identical. Figure 4 also demonstrates that individual contributions to observables are in general shifted between two body neutron-neutron (nn) and 3n forces (and maybe high-body forces) by changing the RG resolution scale of the Hamiltonian. Consequently only the sum of all contributions is an observable. At low resolution scales, (corresponding to the CEFT calculations), the size of 3n contributions to neutron matter pressure is significantly larger than at higher RG resolution scales (corresponding to the QMC and BHF results). Even after the theoretical uncertainties are taken into account, the 3n force significantly increases both the pressure of neutron matter and the neutron skin thickness in $^{208}$Pb. Thus knowledge of the symmetry energy and the skin thickness in $^{208}$Pb may provide information on the 3n forces.

Neutron skin thickness values discussed in Section III are shown in the left-hand panel inset of Figure 4. The skin thickness deduced from the symmetry energy constraints from nucleus-nucleus collisions using various Skyme forces is plotted as a diamond symbol in the inset [79]. The red star in both panels depicts the weighted average of these values, using the relationship of Typel and Brown [53] between the pressure in neutron matter at $\rho/\rho_0$= 0.625 and the neutron skin for $^{208}$Pb. This average value agrees better with calculations that include 3n forces (right panel), but does not have the accuracy to distinguish between these models.

These models can also be used to predict the correlation between $S_0$ and $L$. Like the Skyme interactions, predictions from the RMF models (solid diamonds) from Ref. [80] are scattered on the plane. On the other hand, both the QMC and CEFT models predict a linear dependence of $S_0$ and L as shown by the green dash and blue solid lines in Figure 3, respectively. Both lines are nearly the same and lie below the lower boundary of the experimentally preferred $S_0$ - $L$ correlation, intersecting it only at higher $S_0$ >34 MeV. Figure 3 also shows the $S_0$ - $L$ correlation predicted by BHF [19] and DBHF [20] (squares). Many of the predicted values for L lie near the lower boundary of the experimentally preferred $S_0$ - $L$ correlation; consistent with the results in Fig. 4, where the calculations lie lower than the experimental measurements of the skin thickness of $^{208}$Pb. If future measurements do not shift the $S_0$ - $L$ correlation to lower $L$ some increase in the repulsion of the 3n force maybe required.

## IV.  Summary and outlook

In this article, we have summarized the current status of experimental constraints on the symmetry energy below saturation density, its slope at the saturation density and on the neutron skin thickness of $^{208}$Pb. We have compared results from diverse experiments on a common ground. There is a promising consensus from various experiments using different experimental probes that shows the acceptable range of values of the symmetry energy and its slope to be centered around ($S_O$, L)~(32.5, 70) MeV. The current constraints are applicable to subnormal density, ($0.3\rho_0 \leq \rho \leq \rho_0$) only and are somewhat dependent on the theoretical models used in the analysis of the experiments.

These values are consistent with a neutron skin thickness for $^{208}$Pb of $\delta R_{np}$=0.18±0.027 fm. The skin thickness extracted in PREXII experiment [68] should have much smaller theoretical uncertainties. After $^{208}$Pb measurement, an additional measurement in $^{48}$Ca would be very attractive because microscopic coupled cluster or no-core shell model calculations can closely relate three neutron forces to the skin thickness in $^{48}$Ca [81].

Finally, the density dependence of the symmetry energy has wide ranging ramifications in many branches of nuclear physics and astrophysics, motivating serious efforts to constrain it. Microscopic calculations show that this density dependence depends on poorly constrained 3-neutron forces. Further refinements of theory and experiment and the extension of experimental data to higher and lower densities, particularly relevant for astrophysics, are needed. We note that consistent results have been obtained from both nuclear structure measurements and from the measurement of nucleus-nucleus collisions. These provide support for measurements with high energy nucleus-nucleus collisions [31, 82] designed to probe the symmetry energy at densities of $\rho \approx 2\rho_0$ region.

## V. ACKNOWLEDGEMENTS

The authors thank the organizers and participants of the "International Symposium on Nuclear Symmetry Energy, 2011 (Nusym11)" at Smith College where this project was conceived. The work is supported by the National Science foundation, the Department of Energy and the Robert A. Welch Foundation, LANL LDRD and Open Supercomputing in the United States, the CPAN-MICINN in Spain, Fundação para a Ciência e a Tecnologia (FCT) in Portugal, and the Japan Society for the Promotion of Science.

**References**

[1] The Frontiers of Nuclear Science, Nuclear Science Advisory Committee's December, 2007 Long Range Plan, http://science.energy.gov/np/nsac/
[2] The Nuclear Physics European Collaboration Committee (NuPECC) Long Rang Plan, 2010, http://www.nupecc.org/index.php?display=lrp2010/main
[3] Lattimer, J.M. et al, Science 23, 536-542, (2004).
[4]. The LIGO Scientific Collaboration, Nature Physics 7, 962 (2011).
[5] P. Danielewicz, R. Lacey, W.G. Lynch, Science 298, 1592 (2002).
[6] B. A. Brown, Phys. Rev. Lett. 85, 5296 (2000).
[7] C.J.Horowitz and J.Piekarewicz, Phys. Rev. Lett. **86**, 5647 (2001).
[8] H. A. Bethe and R.F. Bacher, Rev. Mod. Phys. 8 (1936) 82.
[9] C.F. von Weizäcker, Z. Fur Physik, 96, 431 (1935).
[10] P.Moller, W.D.Myers, H.Sagawa and S.Yoshida, Phys. Rev. Lett. 108, 052501 (2012)
[11] D.Vaitherin and D.M.Brink, Phys.Rev.C5, 626 (1972)
[12] J.R.Stone and P.-G.Reinhard, Prog.Part.Nucl.Phys. 58, 587 (2007)
[13] J. Decharge and D. Gogny, Phys. Rev. 21, 1568 (1980).
[14] B. D. Serot and J. D. Walecka, Adv. Nucl. Phys. 16, 1 (1986); Int. J. Mod. Phys. E 6, 515 (1997).
[15] J. Meng et al., Prog. Part. Nucl. Phys 57, 470 (2006)
[16] M. Dutra, O. Lourenco, J. S. Sa Martins, A. Delfino, J.R.Stone and P.D.Steveson, Phys. Rev. C **85**, 035201 (2012)
[17] Steven C. Pieper, V. R. Pandharipande, R. B. Wiringa, and J. Carlson, Phys. Rev. C 64, 014001 (2001)
[18] Steven C. Pieper, Nuclear Physics A 751, 516 (2005)
[19] I.Vidana, C.Providencia, A.Polls and A.Rios, Phys.Rev C80, 045806 (2009)
[20] F.Sammarruca, arXiv:1111.0695; Int. J. Phys. E, Vol.19, 1259 (2010)
[21] S. Gandolfi, A. Yu Illarionov, S. Fantoni, J.C. Miller, F. Pederiva and K.E. Schmidt, Mon. Not. R. Astron. Soc. L35 (2010)
[22] S. Gandolfi, J. Carlson, Sanjay Reddy, Phys. Rev. C 85, 032801(R) (2012).
[23] S. K. Bogner, R. J. Furnstahl, and A. Schwenk, Prog. Part.Nucl. Phys. 65, 94 (2010).
[24] K. Hebeler and A.Schwenk, Phys.Rev. C82, 014314 (2010)
[25] K. Hebeler, J. M. Lattimer, C. J. Pethick, and A. Schwenk, Phys.Rev.Lett. 105, 161102 (2010)
[26] M. B. Tsang, Yingxun Zhang, P. Danielewicz, M. Famiano, Zhuxia Li,
W. G. Lynch and A.W. Steiner, Phys.Rev.Lett. 102, 122701 (2009).
[27] W.G. Lynch, M.B. Tsang, Y. Zhanga, P. Danielewicz, M. Famiano, Z. Li, A.W. Steiner, Prog. Part. Nucl. Phys. 62, 427 (2009).
[28] H.S. Xu, M.B. Tsang, T.X. Liu, X.D. Liu, W.G. Lynch, W.P.Tan and G. Verde, L. Beaulieu, B. Davin, Y. Larochelle, T. Lefort, R.T. de Souza, R. Yanez, V.E. Viola, R.J. Charity, and L.G. Sobotka, Phys. Rev. Lett. **85**, 716 (2000).
[29] M. B. Tsang et al., Phys. Rev. Lett. 92, 062701 (2004).
[30] M. Famiano et al., Phys. Rev. Lett. 97, 052701 (2006)
[31] Russotto et al., Phys. Lett. B697 (2011) 471.
[32] Z.Kohley et al., Phys. Rev. C 82, 064601 (2010);

[33] Z.Kohley et al., Phys. Rev. C 83, 044601 (2011)
[34] M.B. Tsang, C.K. Gelbke, X.D. Liu, W.G. Lynch, W.P. Tan, G. Verde, H.S. Xu, W. A. Friedman, R. Donangelo, S. R. Souza, C.B. Das, S. Das Gupta, and D. Zhabinsky, Phys. Rev. **C64**, 054615 (2001)
[35] T.X. Liu, et al., Phys. Rev. C 76, 034603(2007).
[36] Z. Y. Sun et al., Phys. Rev. C 82, 051603(R) (2010)[
[37] M. S. Antony, A. Pape and J. Britz, Atomic Data and Nuclear Data Tables, 66, 1 (1997).
[38] Pawel Danielewicz and Jenny Lee, AIP Conf. Proc. 1423, 29 (2012).
[39] S. Abrahamyan et al. Phys. Rev. Lett. 108, 112502 (2012).
[40] J. Zenihiro, H. Sakaguchi, T. Murakami, M. Yosoi, Y. Yasuda, S. Terashima, Y. Iwao, H. Takeda, M. Itoh, H. P. Yoshida, and M. Uchida,_ Phys. Rev. C 82, 044611 (2010); J. Zenihiro; PhD Thesis, Kyoto University, 2011].
[41] S. Terashima, H. Sakaguchi, H. Takeda, T. Ishikawa, M. Itoh, T. Kawabata, T. Murakami, M. Uchida, Y. Yasuda, M. Yosoi, J. Zenihiro, H. P. Yoshida, T. Noro, T. Ishida, S. Asaji, and T. Yonemura, Phys.Rev.C**77**, 024317 (2008)
[42] T. Suzuki et al., Phys. Rev. Lett. 75, 3241(1995)
[43] A. Trzcińska, J. Jastrzębski, P. Lubiński, F. J. Hartmann, R. Schmidt, T. von Egidy, and B. Kłos, Phys. Rev. Lett. 87, 082501 (2001)
[44] B. Kłos, A. Trzcińska, J. Jastrzębski, T. Czosnyka, M. Kisieliński, P. Lubiński, P. Napiorkowski, L. Pieńkowski, F. J. Hartmann, B. Ketzer, P. Ring, R. Schmidt, T. von Egidy, R. Smolańczuk, S. Wycech, K. Gulda, W. Kurcewicz, E. Widmann, and B. A. Brown, Phys. Rev. C **76**, 014311 (2007)
[45] M. Centelles, X. Roca-Maza, X. Viñas, M. Warda, Phys. Rev. Lett. 102, 122502 (2009); Phys. Rev. C **80**, 024316 (2009)
[46] A. Tamii et al., PRL 107, 062502 (2011)
[47] Luca Trippa, Gianluca Colo and Enrico Vigezzi, Phys. Rev. C 77, 061304(R) (2008)
[48] A.Klimkiewicz et al., Phys. Rev. C **76**, 051603(R) (2007)
[49] A.Carbone et al., Phys. Rev. C 81, 041301 (2010).
[50] O. Wieland et. al., Phys. Rev. Lett. 102, 092502 (2009).
[51] Bao-An Li, Lie-Wen Chen, Che Ming Ko, Physics Reports 464, 113 (2008)
[52] Y.Z. Zhang et al., Phys. Rev. C, 85, 024602 (2012)
[53] S.Typel and B.A.Brown, Phys.Rev. C 64, 027302 (2001)]
[54] A.W. Steiner, M. Prakash, J.M. Lattimer, P.J. Ellis, Phys. Rep. 411, 325 (2005)
[55] P.Danielewicz and J.Lee, Nucl.Phys. A818, 36 (2009)
[56] C.J. Horowitz, A. Schwenk, Nucl. Phys. A776:55-79,2006
[57] J. B. Natowitz, G. Röpke, S. Typel, D. Blaschke, A. Bonasera, K. Hagel, T. Klähn, S. Kowalski, L. Qin,S. Shlomo, R. Wada and H. H. Wolter, Phys.Rev.Lett. 104, 202501, 2 (2010)
[58] S. R. Souza, M. B. Tsang, B. V. Carlson, R. Donangelo, W. G. Lynch, and A. W. Steiner, Phys. Rev. C 80, 041602(R) (2009) and references therein.
[59] C. Fuchs, Prog. Part. Nucl. Phys. 1, 56 (2006).
[60] J. Rizzo, M. Colonna, V. Baran, M. Di Toro, H.H. Wolter, M. Zielinska-Pfabe, Nucl. Phys. A 806, 79 (2008).
[61] W.D.Myers and W.J.Swiatecki, Annals of Physics, 55,395 (1969): 84, 186 (1974)
[62] P.Danielewicz, Nucl.Phys. A727, 233 (2003)

[63] M. S. Antony, A. Pape and J. Britz, Atomic Data and Nuclear Data Tables, 66, 1 (1997).
[64] G.Fricke et al., At. Data and Nuclear Data Tables, 60, 177 (1995).
[65] B. Alex Brown, G. Shen, G. C. Hillhouse, J. Meng, and A. Trzcińska, Phys. Rev. C **76**, 034305 (2007)
[66] C.J.Horowitz, Phys. Rev. C 57, 3430 (1998); Eur.Phys.J. A30,303 (2006).
[67] C. Horowitz, S.J. Pollock, P.A. Souder, and R. Michaels, Phys. Rev. C 6**3**, 025501 (2001)
[68] P. A. Souder et al., PREX II experimental proposal Jefferson Laboratory (http://hallaweb.jlab.org/parity/prex/prexII.pdf).
[69] N.L. Rodning, L.D. Knutson, W.G. Lynch, and M.B. Tsang; Phys. Rev. Lett. **49** (1982) 909.
[70] B. L. Berman and S. C. Fultz, Rev. Mod. Phys. 47, 713 (1975).
[71] S. Stringari and E. Lipparini, Phys. Lett. B 117,141 (1982).
[73] H. Krivine, J. Treiner and 0. Bohigas, Nucl. Phys. A366, 155 (1980)
[74] J. Piekarewicz, Phys.Rev. C83,034319 (2011)
[75] P.-G.Reinhard and W.Nazarewicz, Phys.Rev. C **81**, 051303(R) (2010)
[76] J.Piekarewicz et al., arXiv:1201.3807
[77] A. W. Steiner and S. Gandolfi, Phys. Rev. Lett. 108, 081102 (2012).
[78] Andrew W. Steiner, James M. Lattimer, and Edward F. Brown, ApJ. 722, 33, (2010)
[79] Jenny Lee, private communications
[80] X. Roca-Maza, M. Centelles, X. Vi˜nas, M. Warda, Phys. Rev. Lett. 106, 252501 (2011)
[81] Shufang Ban, C. J. Horowitz, R. Michaels, J. Phys. G: Nucl. Part. Phys. 39 (2012) 015104.
[82] Symmetry Energy Project, http://groups.nscl.msu.edu/hira/sep.htm

---

[i] For simplicity, we do not include the readily calculated electromagnetic contribution to the energy.

[ii] In the literature, $S_0$ is sometimes also denoted as J.